\documentclass[a4paper,12pt]{article}
\usepackage{amsfonts}
\usepackage{graphicx}

\input{tcilatex}
\begin{document}

\setcounter{page}{0} \topmargin 0pt \oddsidemargin 5mm \renewcommand{%
\thefootnote}{\fnsymbol{footnote}} \newpage \setcounter{page}{0} 
\begin{titlepage}
\begin{flushright}
Berlin Sfb288 Preprint  \\
US-FT/23-99\\
hep-th/9912196\\
\end{flushright}
\vspace{0.2cm}
\begin{center}
{\Large {\bf Thermodynamic Bethe Ansatz of the  Homogeneous
Sine-Gordon models} }

\vspace{0.8cm}
{\large  O.A.~Castro-Alvaredo$^\sharp$, A.~Fring$\,^\star$, C.~Korff$\,^\star$, J.L.~Miramontes$^\sharp$}

\vspace{0.2cm}
{\em $^\sharp$Departamento de F\'\i sica de Part\'\i culas, Facultad de F\'\i sica \\
Universidad de Santiago de Compostela\\
E-15706 Santiago de Compostela, Spain\\
\smallskip
$^\star$Institut f\"ur Theoretische Physik, 
Freie Universit\"at Berlin\\ 
Arnimallee 14, D-14195 Berlin, Germany }
\end{center}
\vspace{0.5cm}
 
\renewcommand{\thefootnote}{\arabic{footnote}}
\setcounter{footnote}{0}

\begin{abstract}
We apply the thermodynamic Bethe Ansatz to investigate the high 
energy behaviour of a class of scattering matrices which have recently been 
proposed  to  describe the Homogeneous  sine-Gordon models related to simply laced Lie algebras. 
 A characteristic feature is that some elements
of the suggested  S-matrices are not parity invariant and contain resonance shifts
which allow for the formation of unstable bound states. From the Lagrangian
point of view  these models may be viewed as  integrable perturbations of
WZNW-coset models and in our analysis we recover indeed in the deep 
ultraviolet regime the  effective central charge  related to these cosets, 
supporting therefore the S-matrix proposal. For the $SU(3)_k$-model we present a
detailed numerical analysis of the scaling function which exhibits the well known 
staircase pattern for theories involving resonance parameters, indicating the energy
scales of stable and unstable particles. We demonstrate that, as a consequence of the
interplay between  the mass scale and the resonance parameter, the ultraviolet limit of the
HSG-model may be viewed alternatively as a  massless ultraviolet-infrared-flow between different
conformal cosets. For $k=2$ we recover as a subsystem the flow between the tricritical Ising and the
Ising model.   

\end{abstract}
\vfill{ \hspace*{-9mm}
\begin{tabular}{l}
\rule{6 cm}{0.05 mm}\\
Castro@fpaxp1.usc.es\\
Fring@physik.fu-berlin.de \\
Korff@physik.fu-berlin.de \\
Miramont@fpaxp1.usc.es
\end{tabular}}
\end{titlepage}
\newpage

\section{Introduction}

The thermodynamic Bethe ansatz (TBA)\ is established as an important method
which serves to investigate ``off-shell'' properties of 1+1 dimensional
quantum field theories. Originally formulated in the context of the
non-relativistic Bose gas by Yang and Yang \cite{Yang}, it was extended
thereafter by Zamolodchikov \cite{TBAZam1} to relativistic quantum field
theories whose scattering matrices factorize into two-particle ones. The
latter property is always guaranteed when the quantum field theory in
question is integrable. Provided the S-matrix has been determined in some
way, for instance via the bootstrap program \cite{Boot} or by extrapolating
semi-classical results, the TBA allows to calculate the ground state energy
of the integrable model on an infinite cylinder whose circumference is
identified as compactified space direction. When the circumference is sent
to zero the effective central charge of the conformal field theory (CFT)\
governing the short distance behaviour can be extracted. In the case in
which the massive integrable field theory is obtained from a conformal model
by adding a perturbative term which breaks the conformal symmetry, the TBA
constitutes therefore an important consistency check for the S-matrix.

The main purpose of this manuscript is to apply this technique to a class of
scattering matrices which have recently been proposed \cite{HSGS} to
describe the Homogeneous sine-Gordon models (HSG) \cite{HSG,Park} related to
simply laced Lie algebras. The latter have been constructed as integrable
perturbations of WZNW-coset theories \cite{Witten} of the form $G_{k}/H$,
where $G$ is a compact simple Lie group, $H\subset G$ a maximal abelian
torus and $k>1$ an integer called the ``level''. These models constitute
particular deformations of coset-models \cite{Witten}, where the specific
choice of the groups ensures that these theories possess a mass gap \cite
{HSG2}. The defining action of the HSG-models reads 
\begin{equation}
S_{\text{HSG}}[g]=S_{\text{CFT}}[g]+\frac{m^{2}}{\pi \beta ^{2}}\,\int
d^{2}x\,\,\left\langle \Lambda _{+},g(\vec{x})^{-1}\Lambda _{-}g(\vec{x}%
)\right\rangle \ \,\;.  \label{HSGaction}
\end{equation}
Here $S_{\text{CFT}}$ denotes the coset action, $\left\langle
\,\,,\,\,\right\rangle $ the Killing form of $G$ and $g(\vec{x})$ a group
valued bosonic scalar field. $\Lambda _{\pm }$ are arbitrary semi-simple
elements of the Cartan subalgebra associated with $H$, which have to be
chosen not orthogonal to any root of G and play the role of continuous
vector coupling constants. The latter constraints do not restrict the
parameter choice in the quantum case with regard to the proposed S-matrix
which makes sense for every choice of $\Lambda _{\pm }$. They determine the
mass ratios of the particle spectrum as well as the behaviour of the model
under a parity transformation. The parameters $m$ and $\beta
^{2}=1/k+O(1/k^{2})$ are the bare mass scale and the coupling constant,
respectively. The non-perturbative definition of the theory is achieved by
identifying $\left\langle \Lambda _{+},g(\vec{x})^{-1}\Lambda _{-}g(\vec{x}%
)\right\rangle $ with a matrix element of the WZNW-field $g(\vec{x})$ taken
in the adjoint representation, which is a spinless primary field of the
coset-CFT and in addition exchanging $\beta ^{2}$ by $1/k$ and $m$ by the
renormalised mass \cite{HSG2}. Some of the conformal data of $S_{\text{CFT}%
}[g]$, which are in principle extractable from the TBA analysis are the
Virasoro central charge $c$ of the coset and the conformal dimensions $%
\Delta ,\bar{\Delta}$ of the perturbing operator in the massless limit 
\begin{equation}
c_{G_{k}}=\frac{k\,\dim G}{k+h}-\ell =\frac{k-1}{k+h}h\ell \,,\quad \quad
\quad \quad \Delta =\bar{\Delta}=\frac{h}{k+h}\;.  \label{cdata}
\end{equation}
Here $\ell $ denotes the rank of $G$ and $h$ its Coxeter number. Since we
have $\Delta <1$ for all allowed values of $k$, the perturbation is always
relevant in the sense of renormalisation\footnote{%
We slightly abuse here the notation and use $c_{G_{k}}$ instead of $%
c_{G_{k}/U(1)^{\ell }}$. Since we always encounter these type of coset in
our discussion, we can avoid bulky expressions in this way.}.

The simplest example of a HSG theory is the complex sine-Gordon model \cite
{Park,CSG} associated with the coset $SU(2)_{k}/U(1)$. As we will argue
below, more complicated HSG theories can be viewed as interacting copies of
complex sine-Gordon theories. The classical equations of motion of these
models correspond to non-abelian affine Toda equations \cite{HSG,Nonab},
which are known to be classically integrable and admit soliton solutions.
Identifying these solutions by a Noether charge allows for a semi-classical
approach to the quantum theory by applying the Bohr-Sommerfeld quantization
rule. The integrability on the quantum level was established in \cite{HSG2}
by the construction of non-trivial conserved charges, which suggests the
factorization of the scattering matrix. Based on the assumption that the
semi-classical spectrum is exact, the S-matrix elements have then been
determined in \cite{HSGS} by means of the bootstrap program for HSG-models
related to simply laced Lie algebras.

The proposed scattering matrix consists partially of $\ell $ copies of
minimal $su(k)$-affine Toda field theories (ATFT) \cite{Roland}, whose mass
scales are free parameters. The scattering between solitons belonging to
different copies is described by an S-matrix which violates parity \cite
{HSGS}. These matrices possess resonance poles and the related resonance
parameters which characterize the formation of unstable bound states are up
to free choice. In the TBA-analysis these resonances lead to the ``staircase
patterns'' in the scaling function, which have been observed previously for
similar models \cite{Stair}. However, in comparison with the models studied
so far, the HSG models are distinguished in two aspects. First they break
parity invariance and second some of the resonance poles can be associated
directly to unstable particles via a classical Lagrangian.

One of the main outcomes of our TBA-analysis is that the suggested \cite
{HSGS} scattering matrix leads indeed to the coset central charge (\ref
{cdata}), which gives strong support to the proposal.

In addition, we present a detailed numerical analysis for the $SU(3)$-HSG\
model, but expect that many of our findings for that case are generalizable
to other Lie groups. The presence of two parameters, i.e. the mass scale and
the resonance parameter allow, similar as for staircase models studied
previously, to describe the ultraviolet limit of the HSG-model alternatively
as the flow between different conformal field theories in the ultraviolet
and infrared regime. We find the following massless flow 
\begin{equation}
UV\equiv SU(3)_{k}/U(1)^{2}\leftrightarrow SU(2)_{k}/U(1)\otimes
SU(2)_{k}/U(1)\equiv IR\,\,.  \label{UVIR}
\end{equation}
We also observe the flow $(SU(3)_{k}/U(1)^{2})/(SU(2)_{k}/U(1))\rightarrow
SU(2)_{k}/U(1)$ as a subsystem inside the HSG-model. For $k=2$ this
subsystem describes the flow between the tricritical Ising and the Ising
model previously studied in \cite{triZam}. In terms of the HSG-model we
obtain the following physical picture: The resonance parameter characterizes
the mass scale of the unstable particles. Approaching the extreme
ultraviolet regime from the infrared we pass various regions: At first all
solitons are too heavy to contribute to the off-critical central charge,
then the two copies of the minimal ATFT will set in, leading to the central
charge corresponding to $IR$ in (\ref{UVIR}) and finally the unstable bound
states will start to contribute such that we indeed obtain (\ref{cdata}) as
the ultraviolet central charge of the HSG-model.

The two values of the resonance parameter $0$ and $\infty $ are special,
corresponding in the classical theory to a choice of the vector couplings in
(\ref{HSGaction}) parallel to each other or orthogonal to a simple root,
respectively. In the former case parity is restored on the classical as well
as on the TBA-level and the central charge corresponding to $UV$ in (\ref
{UVIR}) is also recovered, whereas in the latter case the two copies of the
minimal ATFT are decoupled and unstable bound states may not be produced
leading to the central charge $IR$ in (\ref{UVIR}).

Our manuscript is organized as follows: In section 2 we briefly recall the
main features of the two-particle HSG S-matrix elements stating them also
newly in form of an integral representation. In particular, we comment on
the link between unstable particles and resonance poles as well as on the
loss of parity invariance. In section 3 we introduce the TBA equations for a
parity violating system and carry out the ultraviolet limit recovering the
expected coset central charge. In section 4 we present a detailed study for
the $SU(3)_{k}-$HSG model. We discuss the staircase pattern of the scaling
function and illustrate how the UV limit for the HSG-model may be viewed as
the UV-IR flow between different conformal models. We extract the
ultraviolet central charges of the HSG-models. We study separately the case
when parity is restored, derive universal TBA-equations and Y-systems. In
section 5 we present explicit examples for the specific values $%
k=2,3,4,\infty $. Our conclusions are stated in section 6.

\section{The homogeneous sine-Gordon\ S-matrix}

We shall now briefly recall the main features of the proposed HSG scattering
matrix in a form most suitable for our discussion. Labelling the solitons by
two quantum numbers, we take the two-particle scattering matrix between
soliton $(a,i)$ and soliton $(b,j)$, with $1\leq a,b\leq k-1$ and $1\leq
i,j\leq $ $\ell $, as a function of the rapidity difference $\theta $ to be
of the general form $S_{ab}^{ij}(\theta )$. The particular structure of the
conjectured HSG S-matrix makes it suggestive to refer to the lower indices
as main quantum numbers and to the upper ones as colour. In \cite{HSGS} it
was proposed to describe the scattering of solitons which possess the same
colour by the S-matrix of the $\mathbb{Z}_{k}$-Ising model or equivalently the
minimal $su(k)$-ATFT \cite{Roland} 
\begin{eqnarray}
S_{ab}^{ii}(\theta ) &=&(a+b)_{\theta }\,(|a-b|)_{\theta }\prod_{n=1}^{\min
(a,b)-1}(a+b-2n)_{\theta }^{2}\,\,  \label{S} \\
&=&\exp \int \frac{dt}{t}2\cosh \frac{\pi t}{k}\,\left( 2\cosh \frac{\pi t}{k%
}-I\right) _{ab}^{-1}e^{-it\theta }.  \label{Sint}
\end{eqnarray}
Here we have introduced the abbreviation $(x)_{\theta }=\sinh \frac{1}{2}%
(\theta +i\frac{\pi x}{k})/\sinh \frac{1}{2}(\theta -i\frac{\pi x}{k})$ for
the general building blocks and denote the incidence matrix of the $su(k)$%
-Dynkin diagram by $I$. We re-wrote the above S-matrix from the block form (%
\ref{S}) into a form of an integral representation (\ref{Sint}), since the
latter is more convenient with respect to the TBA analysis. This calculation
may be performed by specializing an analysis in \cite{FKS1,FKS2} to the
particular case at hand. The scattering of solitons with different colour
quantum numbers was proposed to be described by 
\begin{eqnarray}
S_{ab}^{ij}(\theta ) &=&(\eta _{ij})^{ab}\prod\limits_{n=0}^{\min
(a,b)-1}(-|a-b|-1-2n)_{\theta +\sigma _{ij}}\,,\quad \quad \quad
K_{ij}^{g}\neq 0,2  \label{ij} \\
&=&(\eta _{ij})^{ab}\exp -\int \frac{dt}{t}\,\left( 2\cosh \frac{\pi t}{k}%
-I\right) _{ab}^{-1}e^{-it(\theta +\sigma _{ij})},\quad K_{ij}^{g}\neq 0,2\,,
\label{ijint}
\end{eqnarray}
with $K^{g}$ denoting the Cartan matrix of the simply laced Lie algebra $g$.
Here the $\eta _{ij}=\eta _{ji}^{*}$ are arbitrary $k$-th roots of $-1$
taken to the power $a$ times $b$ and the shifts in the rapidity variables
are functions of the vector couplings $\sigma _{ij}$, which are
anti-symmetric in the colour values $\sigma _{ij}=-\sigma _{ji}$. Due to the
fact that these shifts are real, the function $S_{ab}^{ij}(\theta )$ for $%
i\neq j$ will have poles beyond the imaginary axis such that the parameters $%
\sigma _{ji}$ characterize resonance poles. An important feature is that (%
\ref{ij}) is not parity invariant, where parity is broken by the phase
factors $\eta $ as well as the shifts $\sigma $. As a consequence, the usual
relations 
\begin{equation}
S_{ab}^{ii}(\theta )=S_{ba}^{ii}(\theta )=\,S_{ab}^{ii}(-\theta
^{*})^{*}\qquad \text{and \qquad }S_{ab}^{ii}(\theta )S_{ab}^{ii}(-\theta
)=1\,  \label{8}
\end{equation}
satisfied by the parity invariant objects (\ref{S}), are replaced by 
\begin{equation}
S_{ab}^{ij}(\theta )=\,S_{ba}^{ji}(-\theta ^{*})^{*}\qquad \text{and \qquad }%
S_{ab}^{ij}(\theta )S_{ba}^{ji}(-\theta )=1  \label{HS}
\end{equation}
for the scattering between solitons with different colour values. Important
to note is that the first equality in (\ref{8}) has no analogue in (\ref{HS}%
). Thus, instead of being real analytic, as $S_{ab}^{ii}(\theta ),$ the
parity violation forces Hermitian analyticity of $S_{ab}^{ij}(\theta )$ for $%
i\neq j$. Anti-particles are constructed in analogy to the ATFT, that is
from the automorphism which leaves the $su(k)$-Dynkin diagram invariant,
such that $\overline{(a,i)}=(k-a,i)$. The colour of a particle and its
anti-particle is identical. The crossing relation of the S-matrix then reads 
\begin{equation}
S_{\bar{a}b}^{ij}(\theta )=S_{(k-a)b}^{ij}(\theta )=S_{ba}^{ji}(i\pi -\theta
)\,\,.  \label{cross}
\end{equation}
For a general and more detailed discussion of these analyticity issues see 
\cite{HERMAN} and references therein.

Analyzing the above S-matrix we have the following picture concerning the
formation of bound states: Two solitons with the same colour value may form
a bound state of the same colour, whilst solitons of different colour with $%
K_{ij}\neq 0,2$, say $(a,i)$ and $(b,j)$, may only form an unstable state,
say $(\tilde{c},\tilde{k})$ whose lifetime and energy scale are
characterized by the parameter $\sigma $ by means of the Breit-Wigner
formula, see e.g. \cite{BW}, in the form 
\begin{eqnarray}
(M_{\tilde{c}}^{\tilde{k}})^{2}-\frac{(\Gamma _{\tilde{c}}^{\tilde{k}})^{2}}{%
4} &=&(M_{a}^{i})^{2}+(M_{b}^{j})^{2}+2M_{a}^{i}M_{b}^{j}\cosh \sigma \cos
\Theta  \label{BW1} \\
M_{\tilde{c}}^{\tilde{k}}\Gamma _{\tilde{c}}^{\tilde{k}}
&=&2M_{a}^{i}M_{b}^{j}\sinh |\sigma |\sin \Theta \,\,,  \label{BW2}
\end{eqnarray}
where the resonance pole in $S_{ab}^{ij}(\theta )$ is situated at $\theta
_{R}=\sigma -i\Theta $ and $\Gamma _{\tilde{c}}^{\tilde{k}}$ denotes the
decay width of the unstable particle with mass $M_{\tilde{c}}^{\tilde{k}}$.
In the case $a=b$ these unstable states can be identified with solitons in
the semi-classical limit \cite{HSGS,HSGsol}. When $\sigma $ becomes zero, (%
\ref{BW2}) shows that the unstable particles become stable, but are still
not at the same footing as the other asymptotically stable particles. They
become virtual states characterized by poles on the imaginary axis beyond
the physical sheet.

How many free parameters do we have in our model? Computing mass shifts from
renormalisation, we only accumulate contributions from intermediate states
having the same colour as the two scattering solitons. Thus, making use of
the well known fact that the masses of the minimal $su(k)$-affine Toda
theory all renormalise with an overall factor \cite{ATFTS}, i.e. for the
solitons $(a,i)$ we have that $\delta M_{a}^{i}/M_{a}^{i}$ equals a constant
for fixed colour value $i$ and all possible values of the main quantum
number $a$, we acquire in principle $\ell $ different mass scales $%
m_{1},\ldots ,m_{\ell }$ in the HSG-model. In addition there are $\ell -1$
independent parameters in the model in form of the possible phase shifts $%
\sigma _{ij}=-\sigma _{ji}$ for each $i,j$ such that $K_{ij}^{g}\neq 0,2$.
This means overall we have $2\ell -1$ independent parameters in the quantum
theory. There is a precise correspondence to the free parameters which one
obtains from the classical point of view. In the latter case we have the $%
2\ell $ independent components of $\Lambda _{\pm }$ at our free disposal.
This number is reduced by $1$ as a result of the symmetry $\Lambda
_{+}\rightarrow c\Lambda _{+}$ and $\Lambda _{-}\rightarrow c^{-1}\Lambda
_{-}$ which introduces an additional dependence as may be seen from the
explicit expressions for the classical mass ratios and the classical
resonance shifts 
\begin{equation}
\frac{m_{i}}{m_{j}}=\frac{M_{a}^{i}}{M_{a}^{j}}=\sqrt{\frac{(\alpha
_{i}\cdot \Lambda _{+})(\alpha _{i}\cdot \Lambda _{-})}{(\alpha _{j}\cdot
\Lambda _{-})(\alpha _{j}\cdot \Lambda _{+})}},\qquad \quad \sigma _{ij}=\ln
 \sqrt{\frac{(\alpha _{i}\cdot \Lambda _{+})(\alpha _{j}\cdot \Lambda
_{-})}{(\alpha _{i}\cdot \Lambda _{-})(\alpha _{j}\cdot \Lambda _{+})}}
 \,.  \label{sig}
\end{equation}
Here the $\alpha _{i}$ are simple roots.

In comparison with other factorizable scattering matrices involving
resonance shifts, studied in the literature so far, the proposed HSG
scattering matrices differ in two aspects. First of all, they are not parity
invariant and second they allow to associate a concrete Lagrangian
description. The latter fact can be used to support the picture outlined for
the full quantum field theory by a semi-classical analysis. In \cite{HSGsol}
the semi-classical mass for the soliton ($a,i$) was found to be 
\begin{equation}
M_{a}^{i}=\frac{m_{i}}{\pi \beta ^{2}}\,\sin \frac{\pi a}{k}\,
\end{equation}
where $\beta $ is a coupling constant and the $m_{i}$ are the $\ell $
different mass scales.

\section{TBA with parity violation and resonances}

In this section we are going to determine the conformal field theory which
governs the UV regime of the quantum field theory associated with the
S-matrix elements (\ref{S}) and (\ref{ij}). According to the defining
relation (\ref{HSGaction}) and the discussion of the previous section, we
expect to recover the WZNW-coset theory with effective central charge (\ref
{cdata}) in the extreme ultraviolet limit. It is a well established fact
that such high energy limits can be performed by means of the TBA. Since up
to now such an analysis has only been carried out for parity invariant
S-matrices, a few comments are due to implement parity violation. The
starting point in the derivation of the key equations are the Bethe ansatz
equations, which are the outcome of dragging one soliton, say of type $%
A=(a,i)$, along the world line. For the time being we do not need the
distinction between the two quantum numbers. On this trip the formal
wave-function of $A$ picks up the corresponding S-matrix element as a phase
factor when meeting another soliton. Due to the parity violation it matters,
whether the soliton is moved clockwise or counter-clockwise along the world
line, such that we end up with two different sets of Bethe Ansatz equations 
\begin{equation}
e^{iLM_{A}\sinh \theta _{A}}\prod\limits_{B\neq A}S_{AB}(\theta _{A}-\theta
_{B})=1\quad \text{and \quad }e^{-iLM_{A}\sinh \theta
_{A}}\prod\limits_{B\neq A}S_{BA}(\theta _{B}-\theta _{A})=1\,,  \label{BA}
\end{equation}
with $L$ denoting the length of the compactified space direction. These two
sets of equations are of course not entirely independent and may be obtained
from each other by complex conjugation with the help of relation (\ref{HS}).
We may carry out the thermodynamic limit of (\ref{BA}) in the usual fashion 
\cite{TBAZam1} and obtain the following sets of non-linear integral
equations 
\begin{eqnarray}
\epsilon _{A}^{+}(\theta )+\sum_{B}\,\Phi _{AB}*L_{B}^{+}(\theta )
&=&r\,M_{A}\cosh \theta \quad  \label{tba} \\
\epsilon _{A}^{-}(\theta )+\sum_{B}\,\Phi _{BA}*L_{B}^{-}(\theta )
&=&r\,M_{A}\cosh \theta \,\,\,.  \label{ptba}
\end{eqnarray}
As usual we let the symbol '$*$' denote the rapidity convolution of two
functions defined by $f*g(\theta ):=\int d\theta ^{\prime }/2\pi \,f(\theta
-\theta ^{\prime })g(\theta ^{\prime })$. Here $r=m_{1}T^{-1}$ is the
inverse temperature times the overall mass scale $m_{1}$ of the lightest
particle and we also re-defined the masses by $M_{a}^{i}\rightarrow
M_{a}^{i}/m_{1}$ keeping, however, the same notation. As very common in
these considerations we also introduced the so-called pseudo-energies $%
\epsilon _{A}^{+}(\theta )=\epsilon _{A}^{-}(-\theta )$ and the related
functions $L_{A}^{\pm }(\theta )=\ln (1+e^{-\epsilon _{A}^{\pm }(\theta )})$%
. The kernels in the integrals carry the information of the dynamical
interaction of the system and are given by 
\begin{equation}
\Phi _{AB}(\theta )=\Phi _{BA}(-\theta )\;=-i\frac{d}{d\theta }\ln
S_{AB}(\theta )\,.  \label{kernel}
\end{equation}
The statistical interaction is chosen to be of fermionic type. Notice that (%
\ref{ptba}) may be obtained from (\ref{tba}) simply by the parity
transformation $\theta \rightarrow -\theta $ and the first equality in (\ref
{kernel}). The main difference of these equations in comparison with the
parity invariant case is that we have lost the usual symmetry of the
pseudo-energies as a function of the rapidities, such that we have now in
general $\epsilon _{A}^{+}(\theta )\neq \epsilon _{A}^{-}(\theta )$. This
symmetry may be recovered by restoring parity.

The scaling function, which can also be interpreted as off-critical Casimir
energy, may be computed similar as in the usual way 
\begin{equation}
c(r)=\frac{3\,r}{\pi ^{2}}\sum_{A}M_{A}\int\limits_{0}^{\infty }d\theta
\,\cosh \theta \,(L_{A}^{-}(\theta )+L_{A}^{+}(\theta ))\,,  \label{scale}
\end{equation}
once the equations (\ref{tba}) have been solved for the $\epsilon _{A}^{\pm
}(\theta )$. Of special interest is the deep UV limit, i.e. $r\rightarrow 0$%
, of this function since then its value coincides with the effective central
charge $c_{\text{eff}}=c-12(\Delta _{0}+\bar{\Delta}_{0})$ of the conformal
model governing the high energy behaviour. Here $c$ is the Virasoro central
charge and $\Delta _{0},\bar{\Delta}_{0}$ are the lowest conformal
dimensions related to the two chiral sectors of the model. Since the
WZNW-coset is unitary, we expect that $\Delta _{0},\bar{\Delta}_{0}=0$ and $%
c_{\text{eff}}=c$. This assumption will turn out to be consistent with the
analytical and numerical results.

\subsection{Ultraviolet central charge for the HSG model}

In this section we follow the usual argumentation of the TBA-analysis which
leads to the effective central charge, paying, however, attention to the
parity violation. We will recover indeed the value in (\ref{cdata}) as the
central charge of the HSG-models. First of all we take the limits $r,\theta
\rightarrow 0$ of (\ref{tba}) and (\ref{ptba}). When we assume that the
kernels $\Phi _{AB}(\theta )$ are strongly peaked$\footnote{%
That this assuption holds for the case at hand is most easily seen by noting
that the logarithmic derivative of a basic building block $(x)_{\theta }$ of
the S-matrix reads 
\[
-i\frac{d}{d\theta }\ln (x)_{\theta }=-\frac{\sin \left( \frac{\pi }{k}%
\,x\right) }{\cosh \theta -\cos \left( \frac{\pi }{k}\,x\right) }%
=-2\sum_{n=1}^{\infty }\sin \left( \frac{\pi }{k}\,x\right) e^{-n|\theta
|}\,. 
\]
From this we can read off directly the decay properties.}$ at $\theta =0$
and develop the characteristic plateaus one observes for the scaling models,
we can take out the $L$-functions from the integral in the equations (\ref
{tba}), (\ref{ptba}) and obtain similar to the usual situation 
\begin{equation}
\epsilon _{A}^{\pm }(0)+\sum_{B}\,N_{AB}L_{B}^{\pm }(0)=0\qquad \quad \text{%
with \quad }N_{AB}=\frac{1}{2\pi }\int\limits_{-\infty }^{\infty }d\theta
\,\,\Phi _{AB}(\theta )\,\,.  \label{consttba}
\end{equation}
Having the resonance parameter $\sigma $ present in our theory we may also
encounter a situation in which $\Phi _{AB}(\theta )$ is peaked at $\theta
=\pm \sigma $. This means in order for (\ref{consttba}) to be valid, we have
to assume $\epsilon _{A}^{\pm }(0)=\epsilon _{A}^{\pm }(\pm \sigma )$ in the
limit $r\rightarrow 0$ in addition to accommodate that situation. For the
last assumption we will not provide a rigorous analytical argument, but will
justify it instead for particular cases from the numerical results (see e.g.
figure 1). Note that in (\ref{consttba}) we have recovered the parity
invariance.

For small values of $r$ we may approximate, in analogy to the parity
invariant situation, $rM_{A}\cosh \theta $ by $r/2M_{A}\exp \theta $, such
that taking the derivative of the relations (\ref{tba}) and (\ref{ptba})
thereafter yields 
\begin{equation}
\frac{\epsilon _{A}^{\pm }(\theta )}{d\theta }+\frac{1}{2\pi }%
\sum_{B}\int\limits_{-\infty }^{\infty }d\theta ^{\prime }\frac{\Phi
_{AB}(\pm \theta \mp \theta ^{\prime })}{1+\exp (\epsilon _{B}^{\pm }(\theta
^{\prime }))}\frac{d\epsilon _{B}^{\pm }(\theta ^{\prime })}{d\theta
^{\prime }}\,\simeq \frac{r}{2}M_{A}\exp \theta \quad .  \label{der}
\end{equation}
The scaling function acquires the form 
\begin{equation}
c(r)\simeq \frac{3\,r}{2\pi ^{2}}\sum_{A}M_{A}\int\limits_{0}^{\infty
}d\theta \,\exp \theta \,(L_{A}^{-}(\theta )+L_{A}^{+}(\theta ))\,,\quad
\quad \text{for }r\simeq 0  \label{appc}
\end{equation}
in this approximation. Replacing in (\ref{appc}) the term $r/2M_{A}\exp
\theta $ by the l.h.s. of (\ref{der}) a few manipulations lead to 
\begin{equation}
\lim_{r\rightarrow 0}c(r)\simeq \frac{3}{2\pi ^{2}}\sum\limits_{p=+,-}%
\sum_{A}\int\limits_{\epsilon _{A}^{p}(0)}^{\epsilon _{A}^{p}(\infty
)}d\epsilon _{A}^{p}\left[ \ln (1+\exp (-\epsilon _{A}^{p}))+\frac{\epsilon
_{A}^{p}}{1+\exp (\epsilon _{A}^{p})}\right] \,\,\,.
\end{equation}
Upon the substitution $y_{A}^{p}=1/(1+\exp (\epsilon _{A}^{p}))$ we obtain
the well known expression for the effective central charge 
\begin{equation}
c_{\text{eff}}=\frac{6}{\pi ^{2}}\sum_{A}\mathcal{L}\left( \frac{1}{1+\exp
(\epsilon _{A}^{\pm }(0))}\right) \,\,.  \label{ceff}
\end{equation}
Here we used the integral representation for Roger's dilogarithm function $%
\mathcal{L}(x)=1/2\int_{0}^{x}dy(\ln y/(y-1)-\ln (1-y)/y)$, and the facts
that $\epsilon _{A}^{+}(0)=\epsilon _{A}^{-}(0)$, $y_{A}^{+}(\infty
)=y_{A}^{-}(\infty )=0$. This means we end up precisely with the same
situation as in the parity invariant case: Determining at first the phases
of the scattering matrices we have to solve the constant TBA-equation (\ref
{consttba}) and may compute the effective central charge in terms of Roger's
dilogarithm thereafter. Notice that in the ultraviolet limit we have
recovered the parity invariance and (\ref{ceff}) holds for all finite values
of the resonance parameter.

For the case at hand we read off from the integral representation of the
scattering matrices 
\begin{equation}
N_{ab}^{ij}=\delta _{ij}\delta
_{ab}-K_{ij}^{g}\,(K^{su(k)})_{\,\,\,\,\,\,\,\,\,ab}^{-1}\,\,.  \label{NN}
\end{equation}
With $N_{ab}^{ij}$ in the form (\ref{NN}) and the identifications $%
Q_{a}^{i}=\prod_{b=1}^{k-1}(1+\exp (-\epsilon _{b}^{i}(0)))^{K_{ab}^{-1}}$
the constant TBA-equations are precisely the equations which occurred before
in the context of the restricted solid-on-solid models \cite{Kirillov,Kuniba}%
. It was noted in there that (\ref{consttba}) may be solved elegantly in
terms of Weyl-characters and the reported effective central charge coincides
indeed with the one for the HSG-models (\ref{cdata}).

It should be noted that we understand the $N$-matrix here as defined in (\ref
{consttba}) and not as the difference between the phases of the S-matrix. In
the latter case we encounter contributions from the non-trivial constant
phase factors $\eta $. Also in that case we may arrive at the same answer by
compensating them with a choice of a non-standard statistical interaction as
outlined in \cite{BF}.

We would like to close this section with a comment which links our analysis
to structures which may be observed directly inside the conformal field
theory. When one carries out a saddle point analysis, see e.g. \cite{Rich},
on a Virasoro character of the general form 
\begin{equation}
\chi (q)=\sum\limits_{\vec{m}=0}^{\infty }\frac{q^{\frac{1}{2}\vec{m}(%
\mathbf{1}-N)\vec{m}^{t}+\vec{m}\cdot \vec{B}}}{(q)_{1}\ldots (q)_{(k-1)%
\text{$\ell $}}}\,\,,  \label{chi}
\end{equation}
with $(q)_{m}=\prod_{k=1}^{m}(1-q^{k})$, one recovers the set of coupled
equations as (\ref{consttba}) and the corresponding effective central charge
is expressible as a sum of Roger's dilogarithms as (\ref{ceff}). Note that
when we choose $g\equiv A_{1}$ the HSG-model reduces to the minimal ATFT and
(\ref{chi}) reduces to the character formulae in \cite{KM}. Thus the
equations (\ref{consttba}) and (\ref{ceff}) constitute an interface between
massive and massless theories, since they may be obtained on one hand in the
ultraviolet limit from a massive model and on the other hand from a limit
inside the conformal field theory. This means we can guess a new form of the
coset character, by substituting (\ref{NN}) into (\ref{chi}). However, since
the specific form of the vector $\vec{B}$ does not enter in this analysis
(it distinguishes the different highest weight representations) more work
needs to be done in order to make this more than a mere conjecture. We leave
this for future investigations.

\section{The SU(3)$_{k}$-HSG model}

We shall now focus our discussion on $G=SU(3)_{k}$. First of all we need to
establish how many free parameters we have at our disposal in this case.
According to the discussion in section 2 we can tune the resonance parameter
and the mass ratio 
\begin{equation}
\sigma :=\sigma _{21}=-\sigma _{12}\quad \text{and}\quad m_{1}/m_{2}\,\,.
\end{equation}

It will also be useful to exploit a symmetry present in the TBA-equations
related to $SU(3)_{k}$ by noting that the parity transformed equations (\ref
{ptba}) turn into the equations (\ref{tba}) when we exchange the masses of
the different types of solitons. This means the system remains invariant
under the simultaneous transformations 
\begin{equation}
\theta \rightarrow -\theta \quad \quad \text{and\qquad }m_{1}/m_{2}%
\rightarrow m_{2}/m_{1}\,\,.  \label{inv}
\end{equation}
For the special case $m_{1}/m_{2}=1$ we deduce therefore that $\epsilon
_{a}^{(1)}(\theta )=\epsilon _{a}^{(2)}(-\theta )$, meaning that a parity
transformation then amounts to an interchange of the colours. Furthermore,
we see from (\ref{ptba}) and the defining relation $\sigma =\sigma
_{21}=-\sigma _{12}$ that changing the sign of the rapidity variable is
equivalent to $\sigma \rightarrow -\sigma $. Therefore, we can restrict
ourselves to the choice $\sigma \geq 0$ without loss of generality.

\subsection{Staircase behaviour of the scaling function}

We will now come to the evaluation of the scaling function (\ref{scale}) for
finite and small scale parameter $r$. To do this we have to solve first the
TBA equations (\ref{tba}) for the pseudo-energies, which up to now has not
been achieved analytically for systems with a non-trivial dynamical
interaction due to the non-linear nature of the integral equations.
Nonetheless, numerically this problem can be controlled relatively well. In
the UV regime for $r\ll 1$ one is in a better position and can set up
approximate TBA equations involving formally massless particles\footnote{%
The concept of massless scattering has been introduced originally in \cite
{triZam} as follows: The on-shell energy of a right and left moving particle
is given by $E_{\pm }=M/2e^{\pm \theta }$ which is formally obtained from
the on-shell energy of a massive particle $E=m\cosh \theta $ by the
replacement $\theta \rightarrow \theta \pm \sigma /2$ and taking the limit $%
m\rightarrow 0,\sigma \rightarrow \infty $ while keeping the expression $%
M=me^{\theta +\sigma /2}$ finite. It is these on-shell energies we will
encounter in our analysis.} for which various approximation schemes have
been developed which depend on the general form of the L-functions. Since
the latter is not known a priori, one may justify ones assumptions in
retrospect by referring to the numerics. In section 5 we present numerical
solutions for the equations (\ref{tba}) for various levels $k$ showing that
the L-functions develop at most two (three if $m_{1}$ and $m_{2}$ are very
different)plateaus in the range $\ln \frac{r}{2}<\theta <\ln \frac{2}{r}$
and then fall off rapidly (see figure 1). This type of behaviour is similar
to the one known from minimal ATFT \cite{TBAZam1,TBAKM}, and we can
therefore adopt various arguments presented in that context. The main
difficulty we have to deal with here is to find the appropriate ``massless''
TBA equations accommodating the dependence of the TBA equations on the
resonance shifts $\sigma $.

We start by separating the kernel (\ref{kernel}) into two parts 
\begin{eqnarray}
\phi _{ab}(\theta ) &=&\Phi _{ab}^{ii}(\theta )=\int dt\,\left[ \delta
_{ab}-2\cosh \tfrac{\pi t}{k}\left( 2\cosh \tfrac{\pi t}{k}-I\right)
_{ab}^{-1}\right] \,e^{-it\theta }\;,\quad  \label{kernelint} \\
\psi _{ab}(\theta ) &=&\Phi _{ab}^{ij}(\theta +\sigma _{ji})=\int dt\,\left(
2\cosh \tfrac{\pi t}{k}-I\right) _{ab}^{-1}\,e^{-it\theta }\;\quad ,i\neq
j\;.  \label{kernelint2}
\end{eqnarray}
Here $\phi _{ab}(\theta )$ is just the TBA kernel of the $su(k)$-minimal
ATFT and in the remaining kernels $\psi _{ab}(\theta )$ we have removed the
resonance shift. Note that $\phi ,\psi $ do not depend on the colour values $%
i,j$ and may therefore be dropped all together in the notation. The integral
representations for these kernels are obtained easily from the expressions (%
\ref{Sint}) and (\ref{ijint}). They are generically valid for all values of
the level $k$. The convolution term in (\ref{tba}) in terms of $\phi ,\psi $
is then re-written as 
\begin{equation}
\sum_{j=1}^{\ell }\sum_{b=1}^{k-1}\Phi _{ab}^{ij}*L_{b}^{j}(\theta
)=\sum_{b=1}^{k-1}\phi _{ab}*L_{b}^{i}(\theta )+\sum\Sb j=1  \\ j\neq i 
\endSb ^{\ell }\sum_{b=1}^{k-1}\psi _{ab}*L_{b}^{j}(\theta -\sigma _{ji})\,.
\label{conv}
\end{equation}
These equations illustrate that whenever we are in a regime in which the
second term in (\ref{conv}) is negligible we are left with $\ell $
non-interacting copies of the $su(k)$-minimal ATFT.

We will now specialize the discussion on the $su(3)_{k}$-case for which we
can eliminate the dependence on $\sigma $ in the second convolution term by
performing the shifts $\theta \rightarrow \theta \pm \sigma /2$ in the TBA
equations. In the UV limit $r\rightarrow 0$ with $\sigma \gg 1$ the shifted
functions can be approximated by the solutions of the following sets of
integral equations 
\begin{eqnarray}
\varepsilon _{a}^{\pm }(\theta )+\sum_{b=1}^{k-1}\phi _{ab}*L_{b}^{\pm
}\left( \theta \right) +\sum_{b=1}^{k-1}\psi _{ab}*L_{b}^{\mp }\left( \theta
\right) &=&r^{\prime }\,M_{a}^{\pm }\,e^{\pm \theta }\quad \,  \label{uvTba}
\\
\hat{\varepsilon}_{a}^{\pm }(\theta )+\sum_{b=1}^{k-1}\phi _{ab}*\hat{L}%
_{b}^{\pm }\left( \theta \right) &=&r^{\prime }\,M_{a}^{\mp }\,\,e^{\pm
\theta }\,,\quad  \label{kinktba}
\end{eqnarray}
where we have introduced the parameter $r^{\prime }=r\,e^{\frac{\sigma }{2}%
}/2$ familiar from the discussion of massless scattering and the masses $%
M_{a}^{+/-}=M_{a}^{(1)/(2)}$. The relationship between the solutions of the
massless system (\ref{uvTba}), (\ref{kinktba}) and those of the original
TBA-equations is given by 
\begin{eqnarray}
\epsilon _{a}^{(1)/(2)}(\theta ) &=&\varepsilon ^{+/-}(\theta \mp \sigma
/2)\qquad \text{for\quad }\ln (r/2)\ll \pm \theta \ll \ln (r/2)+\sigma
\label{e1} \\
\epsilon _{a}^{(1)/(2)}(\theta ) &=&\hat{\varepsilon}^{-/+}(\theta \pm
\sigma /2)\qquad \text{for\quad }\pm \theta \ll \min [\ln (2/r),\ln
(r/2)+\sigma )]  \label{e2}
\end{eqnarray}
where we have assumed $m_{1}=m_{2}$. Similar equations may be written down
for the generic case. To derive (\ref{e2}) we have neglected here the
convolution terms $(\psi _{ab}*L_{b}^{(1)})(\theta +\sigma )$ and $(\psi
_{ab}*L_{b}^{(2)})(\theta -\sigma )$ which appear in the TBA-equations for $%
\epsilon _{a}^{(2)}(\theta )$ and $\epsilon _{a}^{(1)}(\theta )$,
respectively. This is justified by the following argument: For $|\theta |\gg 
$ $\ln 2/r$ the free on-shell energy term is dominant in the TBA equations,
i.e. $\epsilon _{a}^{i}(\theta )\approx r\,M_{i}^{a}\cosh \theta $ and the
functions $L_{a}^{i}(\theta )$ are almost zero. The kernels $\psi _{ab}$ are
centered in a region around the origin\thinspace $\theta =0$ outside of
which they exponentially decrease, see footnote in section 3.1. for this.
This means that the convolution terms in question can be neglected safely if 
$\theta \ll \ln (r/2)+\sigma $ and $\theta \gg \ln (2/r)-\sigma $,
respectively. Note that the massless system provides a solution for the
whole range of $\theta $ for non-vanishing L-function only if the ranges of
validity in (\ref{e1}) and (\ref{e2}) overlap, i.e. if \quad $\ln (r/2)\ll
\min [\ln (2/r),\ln (r/2)+\sigma ]$ which is always true in the limit $%
r\rightarrow 0$ when $\sigma \gg 0$. The solution is uniquely defined in the
overlapping region. These observations are confirmed by our numerical
analysis below.

The resulting equations (\ref{kinktba}) are therefore decoupled and we can
determine $\hat{L}^{+}$ and $\hat{L}^{-}$ individually. In contrast, the
equations (\ref{uvTba}) for $L_{a}^{\pm }$ are still coupled to each other
due to the presence of the resonance shift. Formally, the on-shell energies
for massive particles have been replaced by on-shell energies for massless
particles in the sense of \cite{triZam}, such that if we interpret $%
r^{\prime }$ as an independent new scale parameter the sets of equations (%
\ref{uvTba}) and (\ref{kinktba}) could be identified as massless TBA systems
in their own right.

Introducing then the scaling function associated with the system (\ref{uvTba}%
) as 
\begin{equation}
c_{\text{o}}(r^{\prime })=\frac{3\,r^{\prime }}{\pi ^{2}}\sum_{a=1}^{k-1}%
\int d\theta \,\,\left[ M_{a}^{+}\,e^{\theta }L_{a}^{+}(\theta
)+\,M_{a}^{-}\,e^{-\theta }L_{a}^{-}(\theta )\right]   \label{c0}
\end{equation}
and analogously the scaling function associated with (\ref{kinktba}) as 
\begin{equation}
\hat{c}_{\text{o}}(r^{\prime })=\frac{3\,r^{\prime }}{\pi ^{2}}%
\sum_{a=1}^{k-1}\int d\theta \,\left[ M_{a}^{+}\,e^{\theta }\hat{L}%
_{a}^{+}(\theta )+M_{a}^{-}\,e^{-\theta }\hat{L}_{a}^{-}(\theta )\right] \;
\label{ckink}
\end{equation}
we can express the scaling function (\ref{scale}) of the HSG model in the
regime $r\rightarrow 0,\;\sigma \gg 1$ approximately by 
\begin{eqnarray}
c(r,\sigma ) &=&\frac{3\,r\,e^{\frac{\sigma }{2}}}{2\pi ^{2}}%
\sum_{i=1,2}\sum_{a=1}^{k-1}M_{a}^{i}\int d\theta \,\left[ \,e^{\theta
}L_{a}^{i}(\theta -\sigma /2)+e^{-\theta }L_{a}^{i}(\theta +\sigma
/2)\right]   \nonumber \\
&\approx &c_{\text{o}}\left( r^{\prime }\right) +\hat{c}_{\text{o}}\left(
r^{\prime }\right) \,\;.  \label{uvscale}
\end{eqnarray}
Thus, we have formally decomposed the massive $SU(3)_{k}$-HSG model in the
UV regime into two massless TBA systems (\ref{uvTba}) and (\ref{kinktba}),
reducing therefore the problem of calculating the scaling function of the
HSG model in the UV limit $r\rightarrow 0$ to the problem of evaluating the
scaling functions (\ref{c0}) and (\ref{ckink}) for the scale parameter $%
r^{\prime }$. The latter depends on the relative size of $\ln (2/r)$ and the
resonance shift $\sigma /2.$ Keeping now $\sigma \gg 0$ fixed, and letting $r
$ vary from finite values to the deep UV regime, i.e. $r=0$, the scale
parameter $r^{\prime }$ governing the massless TBA systems will pass
different regions. For the regime $\ln (2/r)<\sigma /2$ we see that the
scaling functions (\ref{c0}) and (\ref{ckink}) are evaluated at $r^{\prime
}>1$, whereas for $\ln (2/r)>\sigma /2$ they are taken at $r^{\prime }<1$.
Thus, when performing the UV limit of the HSG model the massless TBA systems
pass formally from the ``infrared'' to the ``ultraviolet'' regime with
respect to the parameter $r^{\prime }$. We emphasize that the scaling
parameter $r^{\prime }$ has only a formal meaning and that the physical
relevant limit we consider is still the UV limit $r\rightarrow 0$ of the HSG
model. However, proceeding this way has the advantage that we can treat the
scaling function of the HSG model by the UV and IR central charges of the
systems (\ref{uvTba}) and (\ref{kinktba}) as functions of $r^{\prime }$ 
\begin{equation}
c(r,\sigma )\approx c_{\text{o}}\left( r^{\prime }\right) +\hat{c}_{\text{o}%
}\left( r^{\prime }\right) \approx \left\{ 
\begin{array}{ll}
\,c_{IR}+\hat{c}_{IR}\,, & 0\ll \ln \frac{2}{r}\ll \frac{\sigma }{2} \\ 
c_{UV}+\hat{c}_{UV}\,, & \frac{\sigma }{2}\ll \ln \frac{2}{r}
\end{array}
\right. \,\,.  \label{step}
\end{equation}
Here we defined the quantities $c_{IR}:=\lim_{r^{\prime }\rightarrow \infty
}c_{\text{o}}(r^{\prime })$, $c_{UV}:=\lim_{r^{\prime }\rightarrow 0}c_{%
\text{o}}(r^{\prime })$ and $\hat{c}_{IR},\hat{c}_{UV}$ analogously in terms
of $\hat{c}_{\text{o}}(r^{\prime })$.

In the case of $c_{IR}+\hat{c}_{IR}\neq c_{UV}+\hat{c}_{UV}$, we infer from (%
\ref{step}) that the scaling function develops at least two plateaus at
different heights. A similar phenomenon was previously observed for theories
discussed in \cite{Stair}, where infinitely many plateaus occurred which
prompted to call them ``staircase models''. As a difference, however, the
TBA equations related to these models do not break parity. In the next
subsection we determine the central charges in (\ref{step}) by means of
standard TBA central charge calculation, setting up the so-called constant
TBA equations.

\subsection{Central charges from constant TBA equations}

In this subsection we will perform the limits $r^{\prime }\rightarrow 0$ and 
$r^{\prime }\rightarrow \infty $ for the massless systems (\ref{uvTba}) and (%
\ref{kinktba}) referring to them formally as ``UV-'' and ``IR-limit'',
respectively, keeping however in mind that both limits are still linked to
the UV limit of the HSG model in the scale parameter $r$ as discussed in the
preceding subsection. We commence with the discussion of the ``UV limit'' $%
r^{\prime }\rightarrow 0$ for the subsystem (\ref{uvTba}). We then have
three different rapidity regions in which the pseudo-energies are
approximately given by 
\begin{equation}
\varepsilon _{a}^{\pm }(\theta )\approx \left\{ 
\begin{array}{ll}
r^{\prime }M_{a}\,e^{\pm \theta }, & \text{for }\pm \theta \gg -\ln
r^{\prime } \\ 
-\sum_{b}\phi _{ab}*L_{b}^{\pm }(\theta )-\sum_{b}\psi _{ab}*L_{b}^{\mp
}(\theta ), & \text{for }\ln r^{\prime }\ll \theta \ll -\ln r^{\prime } \\ 
-\sum_{b}\phi _{ab}*L_{b}^{\pm }(\theta ), & \text{for }\pm \theta \ll \ln
r^{\prime }
\end{array}
\right. \;.
\end{equation}
We have only kept here the dominant terms up to exponentially small
corrections. We proceed analogously to the discussion as may be found in 
\cite{TBAZam1,TBAKM}. We see that in the first region the particles become
asymptotically free. For the other two regions the TBA equations can be
solved by assuming the L-functions to be constant. Exploiting once more that
the TBA kernels are centered at the origin and decay exponentially, we can
similar as in section 3.1 take the L-functions outside of the integrals and
end up with the sets of equations 
\begin{eqnarray}
x_{a}^{\pm } &=&\prod_{b=1}^{k-1}(1+x_{b}^{\pm })^{\hat{N}%
_{ab}}(1+x_{b}^{\mp })^{N_{ab}^{\prime }}\quad \quad \text{for }\ln
r^{\prime }\ll \theta \ll -\ln r^{\prime }  \label{ctba1} \\
\hat{x}_{a} &=&\prod_{b=1}^{k-1}(1+\hat{x}_{b})^{\hat{N}_{ab}}\qquad \qquad
\qquad \quad \,\text{for }\pm \theta \ll \ln r^{\prime }  \label{ctba2}
\end{eqnarray}
for the constants $x_{a}^{\pm }=e^{-\varepsilon _{a}^{\pm }(0)}$ and $\hat{x}%
_{a}=e^{-\varepsilon _{a}^{\pm }(\mp \infty )}$. The N-matrices can be read
off directly from the integral representations (\ref{kernelint}) and (\ref
{kernelint2}) 
\begin{equation}
\quad \hat{N}:=\frac{1}{2\pi }\int \phi =1-2(K^{su(k)})^{-1}\qquad \text{%
and\qquad }\;N^{\prime }:=\frac{1}{2\pi }\int \psi =(K^{su(k)})^{-1}\,\,.
\end{equation}
Since the set of equations (\ref{ctba2}) has already been stated in the
context of minimal ATFT and its solutions may be found in \cite{TBAKM}, we
only need to investigate the equations (\ref{ctba1}). These equations are
simplified by the following observation. Sending $\theta $ to $-\theta $ the
constant L-functions must obey the same constant TBA equation (\ref{ctba1})
but with the role of $L_{a}^{+}$ and $L_{a}^{-}$ interchanged. The
difference in the masses $m_{1},m_{2}$ has no effect as long as $m_{1}\sim
m_{2}$ since the on-shell energies are negligible in the central region $\ln
r^{\prime }\ll \theta \ll -\ln r^{\prime }$. Thus, we deduce $%
x_{a}^{+}=x_{a}^{-}=:x_{a}$ and (\ref{ctba1}) reduces to 
\begin{equation}
x_{a}=\prod_{b=1}^{k-1}(1+x_{b})^{N_{ab}}\;\qquad \quad \text{with}\quad
N=1-(K^{su(k)})^{-1}\;.  \label{cTba2a}
\end{equation}
Remarkably, also these set of equations\ may be found in the literature in
the context of the restricted solid-on-solid models \cite{Kuniba}.
Specializing some of the general Weyl-character formulae in \cite{Kuniba} to
the $su(3)_{k}$-case a straightforward calculation leads to 
\begin{equation}
x_{a}=\frac{\sin \left( \frac{\pi \,(a+1)}{k+3}\right) \sin \left( \frac{\pi
\,(a+2)}{k+3}\right) }{\sin \left( \frac{\pi \,\,a}{k+3}\right) \sin \left( 
\frac{\pi \,(a+3)}{k+3}\right) }-1\quad \text{and\quad }\hat{x}_{a}=\frac{%
\sin ^{2}\left( \frac{\pi \,(a+1)}{k+2}\right) }{\sin \left( \frac{\pi \,\,a%
}{k+2}\right) \sin \left( \frac{\pi \,(a+2)}{k+2}\right) }-1\,\,\text{.}
\label{cccTBA}
\end{equation}
Having determined the solutions of the constant TBA equations (\ref{ctba1})
and (\ref{cTba2a}) one can proceed via the standard TBA calculations along
the lines of \cite{TBAZam1,triZam,TBAKM} and compute the central charges
from (\ref{c0}), (\ref{ckink}) and express them in terms of Roger's
dilogarithm function 
\begin{eqnarray}
c_{UV} &=&\lim_{r^{\prime }\rightarrow 0}c_{\text{o}}(r^{\prime })=\frac{6}{%
\pi ^{2}}\sum_{a=1}^{k-1}\left[ 2\mathcal{L}\left( \frac{x_{a}}{1+x_{a}}%
\right) -\mathcal{L}\left( \frac{\hat{x}_{a}}{1+\hat{x}_{a}}\right) \right]
\;,  \label{cuv} \\
\hat{c}_{UV} &=&\lim_{r^{\prime }\rightarrow 0}\hat{c}_{\text{o}}(r^{\prime
})=\frac{6}{\pi ^{2}}\sum_{a=1}^{k-1}\mathcal{L}\left( \frac{\hat{x}_{a}}{1+%
\hat{x}_{a}}\right) \,\,.
\end{eqnarray}
Using the non-trivial identities 
\begin{equation}
\frac{6}{\pi ^{2}}\sum_{a=1}^{k-1}L\left( \frac{x_{a}}{1+x_{a}}\right) =3\,%
\frac{k-1}{k+3}\quad \text{and}\quad \frac{6}{\pi ^{2}}\sum_{a=1}^{k-1}L%
\left( \frac{\hat{x}_{a}}{1+\hat{x}_{a}}\right) =2\,\frac{k-1}{k+2}\;
\end{equation}
found in \cite{Log} and \cite{Kirillov}, we finally end up with 
\begin{equation}
c_{UV}=\frac{\left( k-1\right) (4k+6)}{\left( k+3\right) \left( k+2\right) }%
\qquad \text{and\qquad }\hat{c}_{UV}=2\,\frac{k-1}{k+2}\,\,.  \label{cuvv}
\end{equation}
For the reasons mentioned above $\hat{c}_{UV}$ coincides with the effective
central charge obtained from $su(k)$ minimal ATFT describing parafermions 
\cite{Witten} in the conformal limit. Notice that $c_{UV}$ corresponds to
the coset $(SU(3)_{k}/U(1)^{2})/(SU(2)_{k}/U(1))$.

The discussion of the infrared limit may be carried out completely analogous
to the one performed for the UV limit. The only difference is that in case
of the system (\ref{uvTba}) the constant TBA equations (\ref{ctba1}) drop
out because in the central region the free energy terms becomes dominant
when $r^{\prime }\rightarrow \infty $. Thus in the infrared regime the
central charges of both systems coincide with $\hat{c}_{UV}$, 
\begin{equation}
c_{IR}=\lim_{r^{\prime }\rightarrow \infty }c_{\text{o}}(r^{\prime })=\hat{c}%
_{IR}=\lim_{r^{\prime }\rightarrow \infty }\hat{c}_{\text{o}}(r^{\prime
})=2\,\frac{k-1}{k+2}\;.  \label{cir}
\end{equation}
In summary, collecting the results (\ref{cuvv}) and (\ref{cir}), we can
express equation (\ref{step}) explicitly in terms of the level $k$, 
\begin{equation}
c(r,M_{\tilde{c}}^{\tilde{k}})\approx \left\{ 
\begin{array}{ll}
4\,\frac{k-1}{k+2}\,\,, & \qquad \text{for\quad }1\ll \frac{2}{r}\ll M_{%
\tilde{c}}^{\tilde{k}} \\ 
6\,\frac{k-1}{k+3}\,, & \qquad \text{for\quad }M_{\tilde{c}}^{\tilde{k}}\ll 
\frac{2}{r}
\end{array}
\right. \,\,.  \label{stepk}
\end{equation}
We have replaced the limits in (\ref{step}) by mass scales in order to
exhibit the underlying physical picture. Here $M_{\tilde{c}}^{\tilde{k}}$ is
the smallest mass of an unstable bound state which may be formed in the
process $(a,i)+(b,j)\rightarrow (\tilde{c},\tilde{k})$ for $K_{ij}^{g}\neq
0,2$. We also used that the Breit-Wigner formula (\ref{BW1}) implies that $%
M_{\tilde{c}}^{\tilde{k}}\sim e^{\sigma /2}$ for large positive $\sigma $.

First one should note that in the deep UV limit we obtain the same effective
central charge as in section 3.1, albeit in a quite different manner. On the
mathematical side this implies some non-trivial identities for Rogers
dilogarithm and on the physical (\ref{stepk}) exhibits a more detailed
behaviour than the analysis in section 3.1. In the first regime the lower
limit indicates the onset of the lightest stable soliton in the two copies
of complex sine-Gordon model. The unstable particles are on an energy scale
much larger than the temperature of the system. Thus, the dynamical
interaction between solitons of different colours is ``frozen'' and we end
up with two copies of the $SU(2)_{k}/U(1)$ coset which do not interact with
each other. As soon as the parameter $r$ reaches the energy scale of the
unstable solitons with mass $M_{\tilde{c}}^{\tilde{k}}$, the solitons of
different colours start to interact, being now enabled to form bound states.
This interaction breaks parity and forces the system to approach the $%
SU(3)_{k}/U(1)^{2}$ coset model with central charge given by the formula in (%
\ref{cdata}) for $G=SU(3)$.

The case when $\sigma $ tends to infinity is special and one needs to pay
attention to the order in which the limits are taken, we have 
\begin{equation}
4\,\frac{k-1}{k+2}=\lim_{r\rightarrow 0}\lim_{\sigma \rightarrow \infty
}c(r,\sigma )\neq \lim_{\sigma \rightarrow \infty }\lim_{r\rightarrow
0}c(r,\sigma )=6\,\frac{k-1}{k+3}\,.
\end{equation}

One might enforce an additional step in the scaling function by exploiting
the fact that the mass ratio $m_{1}/m_{2}$ is not fixed. So it may be chosen
to be very large or very small. This amounts to decouple the TBA systems for
solitons with different colour by shifting one system to the infrared with
respect to the scale parameter $r$. The plateau then has an approximate
width of $\sim \ln |m_{1}/m_{2}|$ (see figure 2). However, as soon as $r$
becomes small enough the picture we discussed for $m_{1}\sim m_{2}$ is
recovered.

\subsection{Restoring parity and eliminating the resonances}

In this subsection we are going to investigate the special limit $\sigma
\rightarrow 0$ which is equivalent to choosing the vector couplings $\Lambda
_{\pm }$ in (\ref{HSGaction}) parallel or anti-parallel. For the classical
theory it was pointed out in \cite{HSG} that only then the equations of
motion are parity invariant. Also the TBA-equations become parity invariant
in the absence of the resonance shifts, albeit the S-matrix still violates
it through the phase factors $\eta $. Since in the UV regime a small
difference in the masses $m_{1}$ and $m_{2}$ does not effect the outcome of
the analysis, we can restrict ourselves to the special situation $m_{1}=m_{2}
$, in which case we obtain two identical copies of the system 
\begin{equation}
\epsilon _{a}(\theta )+\sum_{b=1}^{k-1}(\phi _{ab}+\psi _{ab})*L_{b}(\theta
)=r\,M_{a}\cosh \theta \;.
\end{equation}
Then we have $\epsilon _{a}(\theta )=\epsilon _{a}^{(1)}(\theta )=\epsilon
_{a}^{(2)}(\theta )$, $M_{a}=M_{a}^{(1)}=M_{a}^{(2)}$ and the scaling
function is given by the expression 
\begin{equation}
c(r,\sigma =0)=\frac{6\,r}{\pi ^{2}}\sum_{a=1}^{k-1}M_{a}\int d\theta
\,\,L_{a}(\theta )\cosh \theta \;.
\end{equation}
The factor two in comparison with (\ref{scale}) takes the two copies for $%
i=1,2$ into account. The discussion of the high-energy limit follows the
standard arguments similar to the one of the preceding subsection and as may
be found in \cite{TBAZam1,TBAKM}. Instead of shifting by the resonance
parameter $\sigma $, one now shifts the TBA equations by $\ln r/2$. The
constant TBA equation which determines the UV central charge then just
coincides with (\ref{ctba1}). We therefore obtain 
\begin{equation}
\lim_{r\rightarrow 0}\lim_{\sigma \rightarrow 0}c(r,\sigma )=\frac{12}{\pi
^{2}}\sum_{a=1}^{k-1}L\left( \frac{x_{a}}{1+x_{a}}\right) =6\,\frac{k-1}{k+3}%
\;.
\end{equation}
Thus, again we recover the coset central charge (\ref{cdata}) for $G=SU(3)$,
but this time without breaking parity in the TBA equations. This is in
agreement with the results of section 3.1, which showed that we can obtain
this limit for any finite value of $\sigma $.

\subsection{Universal TBA equations and Y-systems}

Before we turn to the discussion of specific examples for definite values of
the level $k$, we would like to comment that there exists an alternative
formulation of the TBA equations (\ref{tba}) in terms of a single integral
kernel. This variant of the TBA equations is of particular advantage when
one wants to discuss properties of the model and keep the level $k$ generic.
By means of the convolution theorem and the Fourier transforms of the TBA
kernels $\phi $ and $\psi $, which can be read off directly from (\ref
{kernelint}) and (\ref{kernelint2}), one derives the set of integral
equations 
\begin{equation}
\epsilon _{a}^{i}(\theta )+\Omega _{k}*L_{a}^{j}(\theta -\sigma
_{ji})=\sum_{b=1}^{k-1}I_{ab}\,\Omega _{k}*(\epsilon
_{b}^{i}+L_{b}^{i})(\theta )\;.\quad  \label{uni}
\end{equation}
We recall that $I$ denotes the incidence matrix of $su(k)$ and the kernel $%
\Omega _{k}$ is found to be 
\begin{equation}
\Omega _{k}(\theta )=\frac{k/2}{\cosh (k\theta /2)}\;.
\end{equation}
The on-shell energies have dropped out because of the crucial relation \cite
{mass} 
\begin{equation}
\sum_{b=1}^{k-1}I_{ab}M_{b}^{i}=2\cos \tfrac{\pi }{k}\,M_{a}^{i}\;,
\end{equation}
which is a property of the mass spectrum inherited from affine Toda field
theory. Even though the explicit dependence on the scale parameter has been
lost, it is recovered from the asymptotic condition 
\begin{equation}
\epsilon _{a}^{i}(\theta )\stackunder{\theta \rightarrow \pm \infty }{%
\longrightarrow }rM_{a}^{i}\,e^{\pm \theta }\,\,\,.
\end{equation}
The integral kernel present in (\ref{uni}) has now a very simple form and
the $k$ dependence is easily read off.

Closely related to the TBA equations in the form (\ref{uni}) are the
following functional relations also referred to as Y-systems. Using complex
continuation (see e.g. \cite{FKS1} for a similar computation) and defining
the quantity $Y_{a}^{i}(\theta )=\exp (-\epsilon _{a}^{i}(\theta ))$ the
integral equations are replaced by 
\begin{equation}
Y_{a}^{i}(\theta +\tfrac{i\pi }{k})Y_{a}^{i}(\theta -\tfrac{i\pi }{k}%
)=\left[ 1+Y_{a}^{j}(\theta -\sigma _{ji})\right] \prod_{b=1}^{k-1}\left[
1+Y_{b}^{i}(\theta )^{-1}\right] ^{-I_{ab}}\,.  \label{Y}
\end{equation}
The Y-functions are assumed to be well defined on the whole complex rapidity
plane where they give rise to entire functions. These systems are useful in
many aspects, for instance they may be exploited in order to establish
periodicities in the Y-functions, which in turn can be used to provide
approximate analytical solutions of the TBA-equations. The scaling function
can be expanded in integer multiples of the period which is directly linked
to the dimension of the perturbing operator.

Noting that the asymptotic behaviour of the Y-functions is $\lim_{\theta
\rightarrow \infty }Y_{a}^{i}(\theta )\sim e^{-rM_{a}^{i}\cosh \theta }$, we
recover for $\sigma \rightarrow \infty $ the Y-systems of the $su(k)$%
-minimal ATFT derived originally in \cite{TBAZamun}. In this case the
Y-systems were shown to have a period related to the dimension of the
perturbing operator (see (\ref{conj})). We found some explicit periods for
generic values of the resonance parameter $\sigma $ as we discuss in the
next section for some concrete examples.

\section{Explicit examples}

In this section we support our analytical discussion with some numerical
results and in particular justify various assumptions for which we have no
rigorous analytical argument so far. We numerically iterate the
TBA-equations (\ref{tba}) and have to choose specific values for the level $%
k $ for this purpose. As we pointed out in the introduction, quantum
integrability has only been established for the choice $k >  h$. Since the
perturbation is relevant also for smaller values of $k$ and in addition the
S-matrix makes perfect sense for these values of $k$, it will be interesting
to see whether the TBA-analysis in the case of $su(3)_{k}$ will exhibit any
qualitative differences for $k \leq 3$ and $k > 3$. From our examples for the
values $k=2,3,4$ the answer to this question is that there is no apparent
difference. For all cases we find the staircase pattern of the scaling
function predicted in the preceding section as the values of $\sigma $ and $%
x $ sweep through the different regimes. Besides presenting numerical plots
we also discuss some peculiarities of the systems at hand. We provide the
massless TBA equations (\ref{uvTba}) with their UV and IR central charges
and state the Y-systems together with their periodicities. Finally, we also
comment on the classical or weak coupling limit $k\rightarrow \infty $.

\subsection{The SU(3)$_{2}$-HSG model}

This is the simplest model for the $su(3)_{k}$ case, since it contains only
the two self-conjugate solitons (1,1) and (1,2). The formation of stable
particles via fusing is not possible and the only non-trivial S-matrix
elements are those between particles of different colour 
\begin{equation}
S_{11}^{11}=S_{11}^{22}=-1,\quad S_{11}^{12}(\theta -\sigma
)=-S_{11}^{21}(\theta +\sigma )=\tanh \frac{1}{2}\left( \theta -i\frac{\pi }{%
2}\right) \;.  \label{ZamS}
\end{equation}
Here we have chosen $\eta _{12}=-\eta _{21}=i$. One easily convinces oneself
that (\ref{ZamS}) satisfies indeed (\ref{HS}) and (\ref{cross}). This
scattering matrix may be related to various matrices which occurred before
in the literature. First of all when performing the limit $\sigma
\rightarrow \infty $ the scattering involving different colours becomes free
and the systems consists of two free fermions leading to the central charge $%
c=1$. Taking instead the limit $\sigma \rightarrow 0$ the expressions in (%
\ref{ZamS}) coincide precisely with a matrix which describes the scattering
of massless ``Goldstone fermions (Goldstinos)'' discussed in \cite{triZam}.
Apart from the factor $i$, the matrix $S_{11}^{21}(\theta )|_{\sigma =0}$
was also proposed to describe the scattering of a massive particle \cite
{anticross}. Having only one colour available one is not able to set up the
usual crossing and unitarity equations and in \cite{anticross} the authors
therefore resorted to the concept of ``anti-crossing''. As our analysis
shows this may be consistently overcome by breaking the parity invariance.
The TBA-analysis is summarized as follows 
\begin{eqnarray*}
\text{unstable particle formation} &:&\text{\qquad \quad }c_{su(3)_{2}}=%
\frac{6}{5}=c_{UV}+\hat{c}_{UV}=\frac{7}{10}+\frac{1}{2} \\
\text{no unstable particle formation} &:&\text{\qquad \quad }%
2c_{su(2)_{2}}=1=c_{IR}+\hat{c}_{IR}=\frac{1}{2}+\frac{1}{2}\,\,\,.
\end{eqnarray*}
It is interesting to note that the flow from the tricritical Ising to the
Ising model which was investigated in \cite{triZam}, emerges as a subsystem
in the HSG-model in the form $c_{UV}\rightarrow c_{IR}$. This suggests that
we could alternatively also view the HSG-system as consisting out of a
massive and a massless fermion, where the former is described by (\ref{c0}),(%
\ref{uvTba}) and the latter by (\ref{ckink}),(\ref{kinktba}), respectively.

Our numerical investigations of the model match the analytical discussion
and justifies various assumptions in retrospect. Figure 1 exhibits various
plots of the L-functions in the different regimes. We observe that for $\ln
(2/r)<\sigma /2,$ $\sigma \neq 0$ the solutions are symmetric in the
rapidity variable, since the contribution of the $\psi $ kernels responsible
for parity violation is negligible. The solution displayed is just the free
fermion L-function, $L^{i}(\theta )=\ln (1+e^{-rM^{i}\cosh \theta })$.
Approaching more and more the ultraviolet regime, we observe that the
solutions $L^{i}$ cease to be symmetric signaling the violation of parity
invariance. The second plateau is then formed, which will extend beyond $%
\theta =0$ for the deep ultraviolet (see figure 1). The staircase pattern of
the scaling function is displayed in figure 2 for the different cases
discussed in the previous section. We observe always the value $6/5$ in the
deep ultraviolet regime, but depending on the value of the resonance
parameter and the mass ratio it may be reached sooner or later. The plateau
at $1$ corresponds to the situation when the unstable particles can not be
formed yet and we only have two copies of $su(3)_{2}$ which do not interact.
Choosing the mass ratios in the two copies to be very different, we can also
``switch them on'' individually as the plateau at $1/2$ indicates.

\vspace{-0.5cm}
\begin{center}
\includegraphics[width=11cm,height=14cm,angle=-90]{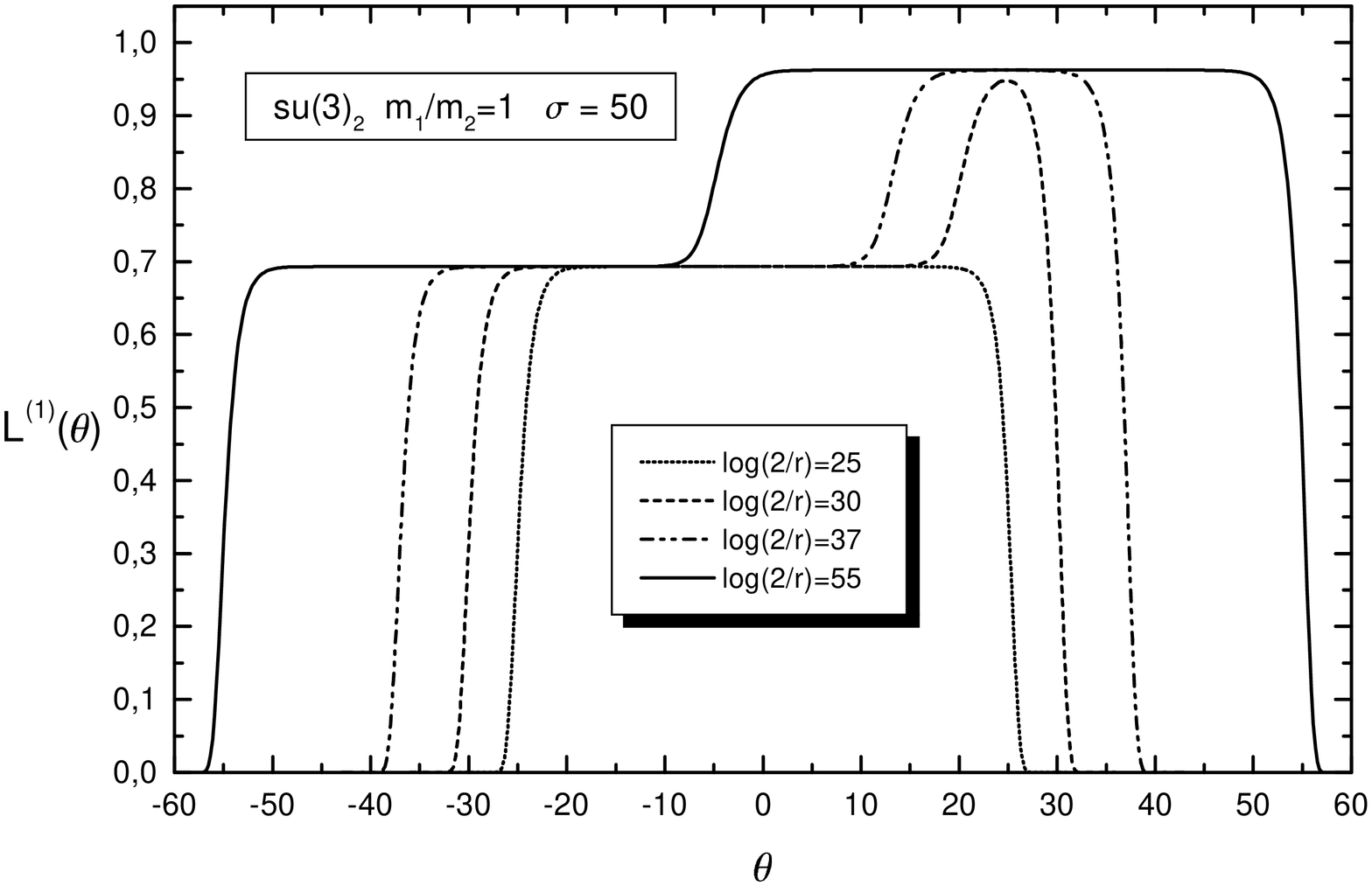}
\end{center}

\vspace*{0.6cm} 
\noindent {\small Figure 1: Numerical solution for $L^{(1)}(\theta )$
of the $su(3)_{2}$ related TBA-equations at different values of the scale
parameter $r$ and fixed resonance shift and mass ratio.}
\vspace*{0.2cm} 

The Y-systems (\ref{Y}) for $k=2$ read 
\begin{equation}
Y_{1}^{i}\left( \theta +i\frac{\pi }{2}\right) Y_{1}^{i}\left( \theta -i%
\frac{\pi }{2}\right) =1+Y_{1}^{j}(\theta -\sigma _{ji})\quad
i,j=1,2,\;i\neq j\,\,.  \label{Y2}
\end{equation}
For $\sigma =0$ they coincide with the ones derived in \cite{triZam} for the
``massless'' subsystem. Shifting the arguments in (\ref{Y2}) appropriately,
the periodicity 
\begin{equation}
Y_{1}^{i}\left( \theta +\frac{5\pi i}{2}+\sigma _{ji}\right)
=Y_{1}^{j}(\theta )\,\,  \label{Periode}
\end{equation}
is obtained after few manipulations. For a vanishing resonance parameter (%
\ref{Periode}) coincides with the one obtained in \cite{TBAZam1,triZam}.
These periods may be exploited in a series expansion of the scaling function
in terms of the conformal dimension of the perturbing operator.

\vspace{-0.2cm}
\begin{center}
\includegraphics[width=11cm,height=16cm,angle=0]{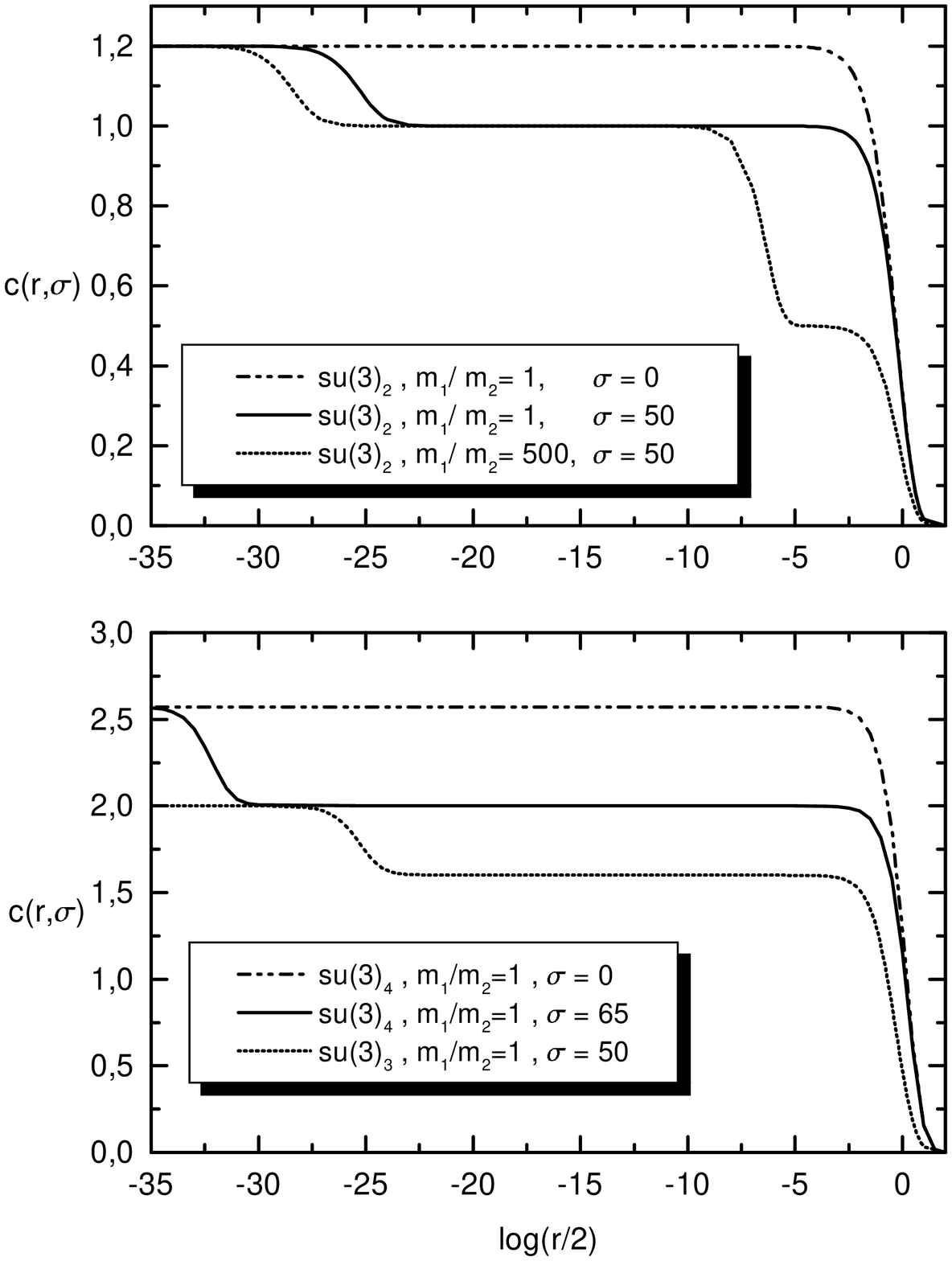}
\end{center}

\vspace*{0.4cm} 
\noindent {\small Figure 2: Numerical plots of the scaling function for 
$su(3)_{k},\;k=2,3,4$ as a function of the variable $\log r/2$ at different
values of the resonance shift and mass ratio.} \vspace*{1.2mm}

\subsection{The SU(3)$_{3}$-HSG model}

This model consists of two pairs of solitons $\overline{(1,1)}=(2,1)$ and $%
\overline{(1,2)}=(2,2)$. When the soliton $(1,i)$ scatters with itself it
may form $(2,i)$ for $i=1,2$ as a bound state. The two-particle S-matrix
elements read 
\begin{equation}
S^{ii}(\theta )=\left( 
\begin{array}{ll}
(2)_{\theta } & -(1)_{\theta } \\ 
-(1)_{\theta } & (2)_{\theta }
\end{array}
\right) \quad \quad S^{ij}(\theta -\sigma _{ij})=\left( 
\begin{array}{ll}
\eta _{ij}\,(-1)_{\theta } & \eta _{ij}^{2}\,(-2)_{\theta } \\ 
\eta _{ij}^{2}\,(-2)_{\theta } & \eta _{ij}\,(-1)_{\theta }
\end{array}
\right) \;.
\end{equation}
Since soliton and anti-soliton of the same colour obey the same TBA
equations we can exploit charge conjugation symmetry to identify $\epsilon
^{i}(\theta ):=\epsilon _{1}^{i}(\theta )=\epsilon _{2}^{i}(\theta )$
leading to the reduced set of equations 
\begin{equation}
\epsilon ^{i}(\theta )+\varphi *L^{i}(\theta )-\varphi *L^{j}(\theta -\sigma
_{ji})=rM^{i}\cosh \theta ,\quad \varphi (\theta )=-\frac{4\sqrt{3}\cosh
\theta }{1+2\cosh 2\theta }\;.
\end{equation}
The corresponding scaling function therefore acquires a factor two, 
\begin{equation}
c(r,\sigma )=\frac{6\,r}{\pi ^{2}}\sum_{i}M^{i}\int d\theta \,\cosh \theta
\,L^{i}(\theta )\;.
\end{equation}
This system exhibits remarkable symmetry properties. We consider first the
situation $\sigma =0$ with $m_{1}=m_{2}$ and note that the system becomes
free in this case 
\begin{equation}
M^{(1)}=M^{(2)}=:M\;\Rightarrow \;\epsilon ^{(1)}(\theta )=\epsilon
^{(2)}(\theta )=rM\cosh \theta \;.
\end{equation}
meaning that the theory falls apart into four free fermions whose central
charges add up to the expected coset central charge of $2$. Also for unequal
masses $m_{1}\neq m_{2}$ the system develops towards the free fermion theory
for high energies when the difference becomes negligible. This is also seen
numerically.

For $\sigma \neq 0$ the two copies of the minimal $A_{2}$-ATFT or
equivalently the scaling Potts model start to interact. The outcome of the
TBA-analysis in that case is summarized as 
\begin{eqnarray*}
\text{unstable particle formation} &:&\text{\qquad \quad }%
c_{su(3)_{3}}=2=c_{UV}+\hat{c}_{UV}=\frac{6}{5}+\frac{4}{5} \\
\text{no unstable particle formation} &:&\text{\qquad \quad }2c_{su(2)_{3}}=%
\frac{8}{5}=c_{IR}+\hat{c}_{IR}=\frac{4}{5}+\frac{4}{5}\,\,\,.
\end{eqnarray*}
As discussed in the previous case for $k=2$ the L-functions develop an
additional plateau after passing the point $\ln (2/r)=\sigma /2$. This
plateau lies at $\ln 2$ which is the free fermion value signaling that the
system contains a free fermion contribution in the UV limit as soon as the
interaction between the solitons of different colours becomes relevant.
Figure 2 exhibits the same behaviour as the previous case, we clearly
observe the plateau at $8/5$ corresponding to the two non-interacting copies
of the minimal $A_{2}$-ATFT. As soon as the energy scale of the unstable
particles is reached the scaling function approaches the correct value of $2$%
.

The Y-systems (\ref{Y}) for $k=3$ read 
\begin{equation}
Y_{1,2}^{i}\left( \theta +i\frac{\pi }{3}\right) Y_{1,2}^{i}\left( \theta -i%
\frac{\pi }{3}\right) =Y_{1,2}^{i}\left( \theta \right) \frac{%
1+Y_{1,2}^{j}(\theta +\sigma _{ij})}{1+Y_{1,2}^{i}\left( \theta \right) }%
\quad i,j=1,2,\;i\neq j\,\,.  \label{Y3}
\end{equation}
Once again we may derive a periodicity 
\begin{equation}
Y_{1,2}^{i}\left( \theta +2\pi i+\sigma _{ji}\right) =Y_{1,2}^{j}(\theta )
\end{equation}
by making the suitable shifts in (\ref{Y3}) and subsequent iteration.

\subsection{The SU(3)$_{4}$-HSG model}

This model involves 6 solitons, two of which are self-conjugate $\overline{%
(2,1)}=(2,1)$, $\overline{(2,2)}=(2,2)$ and two conjugate pairs $\overline{%
(1,1)}=(3,1)$, $\overline{(1,2)}=(3,2)$. The corresponding two-particle
S-matrix elements are obtained from the general formulae (\ref{S})\ and (\ref
{ij}) 
\begin{equation}
S^{ii}(\theta )=\left( 
\begin{array}{ccc}
(2)_{\theta } & (3)_{\theta }(1)_{\theta } & -(2)_{\theta } \\ 
(3)_{\theta }(1)_{\theta } & (2)_{\theta }^{2} & (3)_{\theta }(1)_{\theta }
\\ 
-(2)_{\theta } & (3)_{\theta }(1)_{\theta } & (2)_{\theta }
\end{array}
\right)
\end{equation}
for soliton-soliton scattering with the same colour values and 
\begin{equation}
S^{ij}(\theta -\sigma _{ij})=\left( 
\begin{array}{ccc}
\eta _{ij}(-1)_{\theta } & \eta _{ij}^{2}(-2)_{\theta } & \eta
_{ij}^{3}(-3)_{\theta } \\ 
\eta _{ij}^{2}(-2)_{\theta } & -(-3)_{\theta }(-1)_{\theta } & \eta
_{ij}^{2}(-2)_{\theta } \\ 
\eta _{ij}^{3}(-3)_{\theta } & \eta _{ij}^{2}(-2)_{\theta } & \eta
_{ij}(-1)_{\theta }
\end{array}
\right)
\end{equation}
for the scattering of solitons of different colours with $\eta _{12}=e^{i%
\frac{\pi }{4}}$. In this case the numerics becomes more involved but for
the special case $m_{1}=m_{2}$ one can reduce the set of six coupled
integral equations to only two by exploiting the symmetry $%
L_{a}^{(1)}(\theta )=L_{a}^{(2)}(-\theta )$ and using charge conjugation
symmetry, $L_{1}^{i}(\theta )=L_{3}^{i}(\theta )$. The numerical outcomes,
shown in figure 2 again match, with the analytic expectations (\ref{stepk})
and yield for $\ln (2/r)>\sigma /2$ the coset central charge of $18/7$. In
summary we obtain 
\begin{eqnarray*}
\text{unstable particle formation} &:&\text{\qquad \quad }c_{su(3)_{4}}=%
\frac{18}{7}=c_{UV}+\hat{c}_{UV}=\frac{11}{7}+1 \\
\text{no unstable particle formation} &:&\text{\qquad \quad }%
2c_{su(2)_{4}}=2=c_{IR}+\hat{c}_{IR}=1+1\,\,\,,
\end{eqnarray*}
which matches precisely the numerical outcome in figure 2, with the same
physical interpretation as already provided in the previous two subsections.

\subsection{The semi-classical limit $k\rightarrow \infty $}

As last example we carry out the limit $k\rightarrow \infty $, which is of
special physical interest since it may be identified with the weak coupling
or equivalently the classical limit, as is seen from the relation $\hbar
\beta ^{2}=1/k+O(1/k^{2}).$ To illustrate this equivalence we have
temporarily re-introduced Planck's constant. It is clear from the
TBA-equations that this limit may not be taken in a straightforward manner.
However, we can take it in two steps, first for the on-shell energies and
the kernels and finally for the sum over all particle contributions. The
on-shell energies are easily computed by noting that the mass spectrum
becomes equally spaced for $k\rightarrow \infty $%
\begin{equation}
M_{a}^{i}=M_{k-a}^{i}=\frac{m_{i}}{\pi \beta ^{2}}\sin \frac{\pi \,a}{k}%
\approx a\,m_{i}\;\qquad ,\quad a<\frac{k}{2}\;.  \label{MM}
\end{equation}
For the TBA-kernels the limit may also be taken easily from their integral
representations 
\begin{equation}
\phi _{ab}(\theta )\stackunder{k\rightarrow \infty }{\longrightarrow }2\pi
\,\delta (\theta )\,\left( \delta _{ab}-2\left( K_{ab}^{su(k)}\right)
^{-1}\right) \quad \text{and\quad }\psi _{ab}(\theta )\stackunder{%
k\rightarrow \infty }{\longrightarrow }2\pi \,\delta (\theta )\,\left(
K_{ab}^{su(k)}\right) ^{-1},\;
\end{equation}
when employing the usual integral representation of the delta-function.
Inserting these quantities into the TBA-equations yields 
\begin{equation}
\epsilon _{a}^{i}(\theta )\approx r\,a\,m_{i}\cosh \theta
-\sum_{b=1}^{k-1}\left( \delta _{ab}-2\left( K_{ab}^{su(k)}\right)
^{-1}\right) L_{b}^{i}(\theta )-\sum_{b=1}^{k-1}\left( K_{ab}^{su(k)}\right)
^{-1}L_{b}^{j}(\theta -\sigma )\;.  \label{TTT}
\end{equation}
We now have to solve these equations for the pseudo-energies. In principle
we could proceed in the same way as in the case for finite $k$ by doing the
appropriate shifts in the rapidity. However, we will be content here to
discuss the cases $\sigma \rightarrow 0$ and $\sigma \rightarrow \infty $,
which as follows from our previous discussion correspond to the situation of
restored parity invariance and two non-interacting copies of the minimal
ATFT, respectively. The related constant TBA-equations (\ref{ctba2}) and (%
\ref{cTba2a}) become 
\begin{equation}
\sigma \rightarrow \infty :\;\hat{x}_{a}\stackunder{k\rightarrow \infty }{%
\longrightarrow }\frac{(a+1)^{2}}{a(a+2)}-1\quad \text{and\quad }\sigma
\rightarrow 0:\;x_{a}\stackunder{k\rightarrow \infty }{\longrightarrow }%
\frac{(a+1)(a+2)}{a(a+3)}-1\;.  \label{l1}
\end{equation}
The other information we may exploit about the solutions of (\ref{TTT}) is
that for large rapidities they tend asymptotically to the free solution,
meaning that 
\begin{equation}
\sigma \rightarrow 0,\infty \;:\;\;\;\;\;L_{a}^{i}(\theta )\stackunder{%
\theta \rightarrow \pm \infty }{\longrightarrow }\ln (1+e^{-r\,a\,m_{i}\cosh
\theta })\;.  \label{l2}
\end{equation}
We are left with the task to seek functions which interpolate between the
properties (\ref{l1}) and (\ref{l2}). Inspired by the analysis in \cite
{Fowler} we take these functions to be 
\begin{eqnarray}
\sigma &\rightarrow &\infty \;:\;\;\;\;\;L_{a}^{i}(\theta )=\ln \left[ \frac{%
\sinh ^{2}\left( \frac{a+1}{2}\,rm_{i}\cosh \theta \right) }{\sinh \left( 
\frac{a\,}{2}\,rm_{i}\cosh \theta \right) \sinh \left( \frac{a+2}{2}%
\,rm_{i}\cosh \theta \right) }\right]  \label{p1} \\
\sigma &\rightarrow &0\;:\;\;\;\;\;L_{a}^{i}(\theta )=\ln \left[ \frac{\sinh
\left( \frac{a+1}{2}\,rm_{i}\cosh \theta \right) \sinh \left( \frac{a+2}{2}%
\,rm_{i}\cosh \theta \right) }{\sinh \left( \frac{a}{2}\,rm_{i}\cosh \theta
\right) \sinh \left( \frac{a+3}{2}\,rm_{i}\cosh \theta \right) }\right] \,\,.
\label{p2}
\end{eqnarray}
The expression (\ref{p1}) coincides with the expressions discussed in the
context of the breather spectrum of the sine-Gordon model \cite{Fowler} and (%
\ref{p2}) is constructed in analogy. We are now equipped to compute the
scaling function in the limit $k\rightarrow \infty $%
\begin{equation}
c(r,\sigma )=\lim_{k\rightarrow \infty }\frac{3\,r}{\pi ^{2}}%
\sum_{i=1}^{2}\int d\theta \,\cosh \theta
\sum_{a=1}^{k-1}M_{a}^{i}L_{a}^{i}(\theta )\,\,.
\end{equation}
Using (\ref{MM}), (\ref{p1}) and (\ref{p2}) the sum over the main quantum
number may be computed directly by expanding the logarithm. We obtain for $%
k\rightarrow \infty $%
\begin{eqnarray}
c(r)|_{\sigma =\infty }\!\! &=&\!\!\dfrac{-6r}{\pi ^{2}}\!\!\sum_{i=1}^{2}\!%
\!\int \!\!d\theta \,m_{i}\cosh \theta \ln \left( 1-e^{-r\,m_{i}\cosh \theta
}\right)  \label{in1} \\
c(r)|_{\sigma =0}\!\! &=&\!\!\dfrac{-6\,r}{\pi ^{2}}\!\!\sum_{i=1}^{2}\!\!%
\int \!\!d\theta \,\,m_{i}\cosh \theta [\ln \left( 1-e^{-r\,m_{i}\cosh
\theta }\right) +\ln (1-e^{-r\,2m_{i}\cosh \theta })].\,\,\quad \,
\label{in2}
\end{eqnarray}
Here we have acquired an additional factor of 2, resulting from the
identification of particles and anti-particles which is needed when one
linearizes the masses in (\ref{MM}). Taking now the limit $r\rightarrow 0$
we obtain 
\begin{eqnarray}
\text{no unstable particle formation} &:&\text{\quad \ }2\,c_{su(2)_{\infty
}}=4\;  \label{in11} \\
\text{unstable particle formation} &:&\text{\quad \ }c_{su(3)_{\infty
}}=6\,\,.  \label{in22}
\end{eqnarray}

The results (\ref{in1}), (\ref{in11}) and (\ref{in2}), (\ref{in22}) allow a
nice physical interpretation. We notice that for the case $\sigma
\rightarrow \infty $ we obtain four times the scaling function of a free
boson. This means in the classical limit we obtain twice the contribution of
the non-interacting copies of $SU(2)_{\infty }/U(1)$, whose particle content
reduces to two free bosons each of them contributing $1$ to the effective
central charge which is in agreement with (\ref{cdata}). For the case $%
\sigma \rightarrow 0$ we obtain the same contribution, but in addition the
one from the unstable particles, which are two free bosons of mass $2m_{i}$.
This is also in agreement with (\ref{cdata}).

Finally it is interesting to observe that when taking the resonance poles to
be $\theta _{R}=\sigma -i\pi /k$ the semi-classical limit taken in the
Breit-Wigner formula (\ref{BW1}) leads to $m_{\tilde{k}%
}^{2}=(m_{i}+m_{j})^{2}$. On the other hand (\ref{in2}) seems to suggest
that $m_{\tilde{k}}=2m_{i}$, which implies that the mass scales should be
the same. However, since our analysis is mainly based on exploiting the
asymptotics we have to be cautious about this conclusion.

\section{Conclusions}

Our main conclusion is that the TBA-analysis indeed confirms the consistency
of the scattering matrix proposed in \cite{HSGS}. In the deep ultraviolet
limit we recover the $G_{k}/U(1)^{\ell }$-coset central charge for any value
of the $2\ell -1$ free parameters entering the S-matrix, including the
choice when the resonance parameters vanish and parity invariance is
restored on the level of the TBA-equations. This is in contrast to the
properties of the S-matrix, which is still not parity invariant due to the
occurrence of the phase factors $\eta $, which are required to close the
bootstrap equations \cite{HSGS}. However, they do not contribute to our
TBA-analysis, which means that so far we can not make any definite statement
concerning the necessity of the parity breaking, since the same value for
the central charge is recovered irrespective of the value of the $\sigma $%
's. The underlying physical behaviour is, however, quite different as our
numerical analysis demonstrates. For vanishing resonance parameter the deep
ultraviolet coset central charge is reached straight away, whereas for
non-trivial resonance parameter one passes the different regions in the
energy scale. Also the choice of different mass scales leads to a theory
with a different physical content, but still possessing the same central
charge. To settle this issue, it would therefore be highly desirable to
carry out the series expansion of the scaling function in $r$ and determine
the dimension $\Delta $ of the perturbing operator. It will be useful for
this to know the periodicities of the Y-functions. We conjecture that they
will be

\begin{equation}
Y_{a}^{i}\left( \theta +i\pi (1-\Delta )^{-1}+\sigma _{ji}\right) =Y_{\bar{a}%
}^{j}(\theta ),  \label{conj}
\end{equation}
which is confirmed by our $su(N)$-examples. For vanishing resonance
parameter and the choice $g=su(2)$, this behaviour coincides with the one
obtained in \cite{TBAZamun}. This means the form in (\ref{conj}) is of a
very universal nature beyond the models discussed here.

We also observe from our $su(N)$-example that the different regions, i.e. $%
k > h$ and $k \leq h$, for which quantum integrability was shown and for which
not, respectively, do not show up in our analysis.

It would be very interesting to extend the case-by-case analysis of section
5 to other algebras. The first challenge in these cases is to incorporate
the different resonance parameters.

\medskip

\noindent \textbf{Acknowledgments: } A.F. and C.K. are grateful to the
Deutsche Forschungsgemeinschaft (Sfb288) for financial support and would
also like to thank the Departamento de F\'{\i }sica de Part\'{\i }culas,
Universidad de Santiago de Compostela for kind hospitality. O.A.C. and
J.L.M. thank CICYT (AEN99-0589), DGICYT (PB96-0960), and the EC Commission
(TMR grant FMRX-CT96-0012) for partial financial support. We are grateful to
J. S\'{a}nchez Guill\'{e}n for useful discussions and A. Kuniba for
providing informations concerning one of his preprints.

\newpage

\begin{description}
\item  {\small \setlength{\baselineskip}{12pt}}
\end{description}

\end{document}